\documentclass[aps,pre,twocolumn,showpacs]{revtex4}
\usepackage{graphicx}%
\usepackage{dcolumn}
\usepackage{amsmath}   

\usepackage{bm}
\usepackage{longtable}

\begin{document}

\title{ Electrostatic fluctuations in cavities within polar liquids
  and thermodynamics of polar solvation }
\author{Daniel R.\ Martin}
\author{Dmitry V.\ Matyushov}
\affiliation{Center for Biological Physics, Arizona State University, 
PO Box 871604, Tempe, AZ 85287-1604                   }

\begin{abstract}
  We present the results of numerical simulations of fluctuations of
  the electrostatic potential and electric field inside cavities
  created in the fluid of dipolar hard spheres. We found that the
  thermodynamics of polar solvation dramatically changes its regime
  when the cavity size becomes about 4-5 times larger than the size of
  the liquid particle. The range of small cavities can be reasonably
  understood within the framework of current solvation models. On the
  contrary, the regime of large cavities is characterized by a
  significant softening of the cavity interface resulting in a decay
  of the fluctuation variances with the cavity size much faster than
  anticipated by both the continuum electrostatics and microscopic
  theories. For instance, the variance of potential decays with the
  cavity size $R_0$ approximately as $1/R_0^{4-6}$ instead of the
  $1/R_0$ scaling expected from standard electrostatics. Our results
  suggest that cores of non-polar molecular assemblies in polar
  liquids lose solvation strength much faster than is traditionally
  anticipated.
\end{abstract}

\maketitle

\section{Introduction}
\label{sec:1}
Solvation represents the change in the free energy when a usually
molecular object is inserted into a condensed-phase environment. Since
a significant part of chemistry and all life processes happen in
liquid solutions, the traditional focus has been on solvation in
liquids, polar liquids in particular. The heterogeneous problem of
solvation is probably as complex as the theory of liquids itself and
is hunted by the same basic issues making the quantitative description
of liquids so hard. There are two dominating and mutually compensating
contributions to the free energy of solvation: the positive free
energy of creating a cavity (empty space) for a molecule to be
inserted and a negative stabilization energy from short range (van der
Waals) and long-range (electrostatic) forces \cite{Beck:06}. The
positive cavity free energy is normally significantly compensated by
the negative stabilization free energy resulting in the overall
solvation free energy, a situation akin to the competition between
repulsive and attractive forces in equilibrium liquids
\cite{Hansen:03}.

The present study is devoted to electrostatic solvation, i.e.\ the
free energy arising from the electrostatic interactions between the
charge distribution of the solute with the charge distribution of the
liquid solvent. The charge distribution within molecular solutes is
often modeled by atomic partial charges efficiently used in force
fields of numerical simulations. On the contrary, the charge
distribution of the solvent molecules is often well represented by
molecular multipoles following the well-established tradition of
classical electrostatics \cite{Landau8} and dielectric theory
\cite{Boettcher:73}. Extensions to models utilizing atomic charges are
also possible as used in numerical simulations \cite{Allen:96} and
interaction-site models of molecular liquids \cite{Hansen:03}.

Electrostatic solvation is believed to be well-understood. Following
Born \cite{Born:20} and Onsager \cite{Onsager:36}, the problem is
traditionally recast in terms of continuum electrostatics where the
electrostatic free energy is sought for the solute charges inserted
into a dielectric cavity. This approach has been extensively tested
against the experimental database of solvation of small ions and
neutral molecules in polar molecular liquids
\cite{Marenich:08}. Despite some inconsistencies, the formalism can be
easily incorporated into quantum calculations and can even be
quantitative once the dielectric cavity is properly parametrized.

There are however still some fundamental issues that cannot be
addressed within elecrostatic models. The solution of the Poisson
equation in dielectric media is essentially a boundary condition
problem in which the assumptions tacitly made by the material
Maxwell's equations about the structure of the dielectric interface
are essential for the solution. The standard electrostatics assumes
abrupt discontinuity of the dipolar polarization at the dielectric
surface. This boundary condition creates surface charge
\cite{Boettcher:73} which is ultimately responsible for the
electrostatic potential within the dielectric cavity. Whether
interfaces of real polar liquids \cite{Sokhan:97} match the assumption
of abrupt discontinuity of the bulk polarization is an open
question. For instance, the electric field within a cavity in a polar
liquid was found to be much different from the prediction of standard
electrostatics up to the cavity size of a mesoscale dimension
\cite{DMepl:08}.

A new additional piece of evidence comes from studies of hydrophobic
solvation essential for colloid stability, biopolymer folding, and
formation of biological supramolecular structures
\cite{Rashin:93,ChandlerNature:05}. It was found that solvation of
non-polar solutes changes dramatically in character at the length of
about 1 nm, which is about three molecular diameters for aqueous
solvation \cite{Lum:99}. Solvation of solutes larger than this
characteristic length was found to be dominated by surface effects,
i.e. the structure of water at the hydrophobic interface. Weak
dewetting \cite{HummerPRL:98,Huang:00}, i.e.\ a substantial decrease
of the water density at the interface compared to the bulk water, was
found to be a central part of solvation of large hydrophobic solutes.

Given the current interest in solvation at mesoscale
\cite{Maggs:06,Ashbaugh:06}, to a large extent driven by biological
applications \cite{Rudas:06}, we address here the problem of
electrostatic solvation of solutes significantly larger than have been
mostly studied so far. Our study is driven by the question whether the
change in the solvation character established for hydrophobic solutes
\cite{ChandlerNature:05} is reflected in an equally dramatic change in
the character of electrostatic solvation. The fact that the properties
of a polar liquid interface are inconsistent with the assumptions of
Maxwell's electrostatics \cite{DMepl:08} points to the possibility of
a new solution once the size of the solute exceeds some critical
dimension. This is indeed the result we report here.

We have found from numerical simulations that the scaling of the
fluctuations of the electrostatic potential and electric field with
the cavity radius is consistent with the expectations of
electrostatics (qualitatively) and molecular solvation models
(quantitatively) for small solutes, but changes dramatically at
approximately the same solute/solvent size ratio as observed for
hydrophobic solvation. It turns out that the core of the solute
becomes non-polar with its growing size much faster than is normally
anticipated. We will start with formulating the general results of the
Gaussian solvation thermodynamics and discuss the outcome of computer
simulations next.

\section{Thermodynamics of electrostatic solvation}
\label{sec:2}
By definition, the chemical potential of electrostatic solvation is
given by the ratio of two partition functions: the one which
includes the electrostatic solute-solvent potential $V_{0s}$ and the
one which is based on the non-electrostatic solute-solvent
interactions and the interactions between the solvent particles. All
these latter interactions are incorporated in the Hamiltonian $H_0$.
The relation for $\mu_{0s}$ is then
\begin{equation}
  \label{eq:1}
  e^{-\beta \mu_{0s}(\beta) } = Q(\beta)^{-1} \int e^{-\beta V_{0s} - \beta H_0} d\Gamma ,
\end{equation}
where
\begin{equation}
  \label{eq:2}
  Q(\beta) = \int e^{-\beta H_0} d\Gamma .
\end{equation}
Here, we use the subscript ``0'' for the solute and the subscript
``s'' for the solvent, $d\Gamma$ denotes integration over the system phase
space, and $\beta$ is the inverse temperature. Equation (\ref{eq:1}) can
be conveniently re-written in terms of the product of the Boltzmann distribution 
of finding the solute-solvent energy $\epsilon=V_{0s}$ and the probability 
density $P(\epsilon,\beta)$
\begin{equation}
  \label{eq:3}
  e^{-\beta \mu_{0s}(\beta)} = \int P(\epsilon,\beta) e^{-\beta \epsilon} d\epsilon, 
\end{equation}
where
\begin{equation}
  \label{eq:4}
  P(\epsilon,\beta)= Q(\beta)^{-1}\int \delta(\epsilon - V_{0s}) e^{-\beta H_0} d\Gamma .  
\end{equation}

Equation (\ref{eq:3}) is exact and it states that all the
thermodynamic information required to understand electrostatic
solvation is contained in the distribution of fluctuations of the
interaction energy $\epsilon=V_{0s}$ produced by the solvent which is
actually not polarized by this potential; $V_{0s}=0$ for the
Hamiltonian $H_0$.

The approximation that we will adopt in our formalism, which is
supported by our present simulations and data from other groups
\cite{Kuharski:88,Aqvist:96,Blumberger:05}, is to assume that the
distribution function $P(\epsilon, \beta)$ is a Gaussian function with zero
average
\begin{equation}
  \label{eq:5}
  P(\epsilon,\beta) \propto \exp\left[ - \frac{\epsilon^2}{2\sigma^2(\beta)}\right] .
\end{equation}
The approximation of zero average is the reflection of the fact that
no specific orientation of the solvent dipoles is created around a
non-polar solute. This approximation is not necessarily always correct
\cite{Sokhan:97,Ashbaugh:00,Cerutti:07}, but is insignificant for most
of our development since a non-zero average, if it exists, can always
be incorporated in a linear shift of $\epsilon$. What is the most significant
property for our analysis is the magnitude and the temperature dependence
of the Gaussian width $\sigma^2(\beta)$.

Within the Gaussian approximation for the electrostatic fluctuations
around a non-polar solute the thermodynamics of solvation gains a
simple and physically transparent form. The chemical potential of
solvation is
\begin{equation}
  \label{eq:6}
  \mu_{0s} = -(\beta/2) \sigma^2(\beta) .
\end{equation}
In addition, one can determine the energy $e$ and entropy $s$ of
electrostatic solvation
\begin{equation}
  \label{eq:7}
  \begin{split}
  e & = \langle V_{0s} \rangle + \Delta e_{ss}, \\
  Ts & = \frac{ \langle V_{0s} \rangle}{2} + \Delta e_{ss} .
  \end{split}
\end{equation}
In this equation, $\langle V_{0s} \rangle$ is the average solute-solvent electrostatic
interaction energy when full solute-solvent interaction is turned on.
From Eqs.\ (\ref{eq:6}) and (\ref{eq:7}),
\begin{equation}
  \label{eq:18}
  \langle V_{0s}\rangle  = - \beta \sigma^2(\beta) .
\end{equation}

The term $\Delta e_{ss}$ in Eq.\ (\ref{eq:7}) determines the change in the
interaction energy between the solvent molecules induced by
electrostatic solute-solvent interaction. This energy term is
identically equal to the corresponding contribution to the solvation
entropy, $T\Delta s_{ss}=\Delta e_{ss}$, so that $\Delta e_{ss}$ cancels out in the
solvation chemical potential which is determined by solute-solvent
interaction thermodynamics only \cite{Yu:88,Ben-Amotz:05}.  The term
$\Delta e_{ss}$ can be calculated by either taking the derivative of the
Gaussian width $\sigma^2(\beta)$ or from a third-order correlation function
\begin{equation}
  \label{eq:8}
  \Delta e_{ss} = -\frac{\beta^2}{2} \frac{\partial \sigma^2}{\partial \beta} = (\beta^2/2) \langle \delta
  V_{0s}^2\delta H_0\rangle_0 .
\end{equation}
In Eq.\ (\ref{eq:8}), the average $\langle\dots\rangle_0$ is over the ensemble of
the non-polar solute in equilibrium with the solvent, collectively
described by the Hamiltonian $H_0$. In addition, $\delta V_{0s}= V_{0s} -\langle
V_{0s}\rangle_0 $ and $\delta H_0=H_0 - \langle H_0\rangle_0$ are deviations from the average
values determined on the same unpolarized ensemble.

\section{Simulations and data analysis}
\label{sec:3}
\begin{figure}
  \centering
  \includegraphics*[width=7cm]{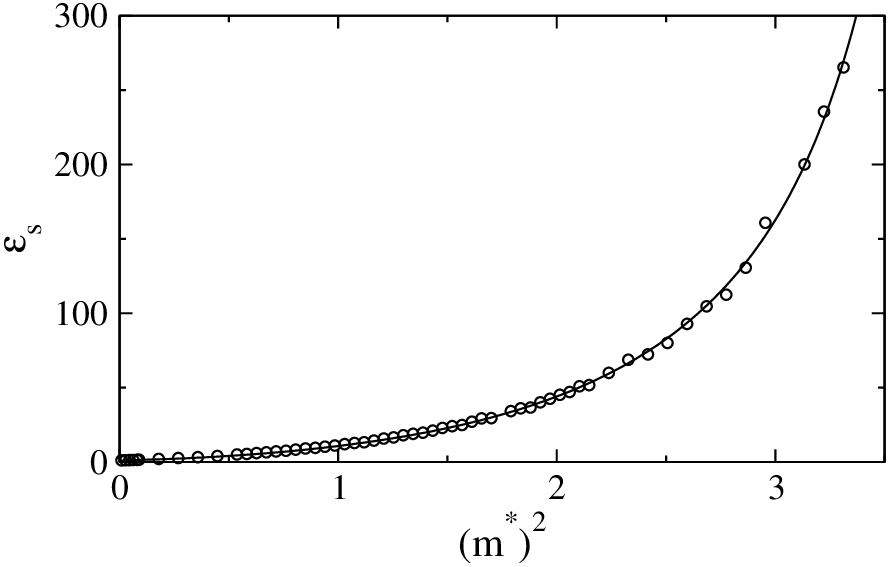}
  \caption{Dielectric constant $\epsilon_s$ of the liquid of dipolar hard
    spheres vs the dipolar parameter $(m^*)^2=\beta m^2/ \sigma^3$; $\rho^*=0.8$.
    The solid line represents the Pad\'e approximation of the simulation
    data: $\epsilon_s(x)=(1+a_1 x + a_2 x^2)/(1+ b_1 x+b_2 x^2)$ with
    $a_1=2.506$, $a_2=3.057$, $b_1=-0.180$, $b_2=-0.00865$ and
    $x=(m^*)^2$.  The dielectric constants were calculated from NVT MC
    simulations of the homogeneous liquid of dipolar hard spheres
    using the Neumann \cite{Neumann:86} correction for the cutoff of
    dipolar interactions treated by the reaction-field formalism.  }
  \label{fig:1}
\end{figure}

While the equations presented in Sec.\ \ref{sec:2} are generally
applicable to an arbitrary solute, we will use numerical Monte Carlo
(MC) simulations \cite{Allen:96} to determine the statistics of
fluctuations produced in spherical cavities carved from a liquid of
dipolar hard spheres (see Appendix for the description of the
simulation protocol). The fluid of dipolar hard spheres leaves out
many important properties of real liquids, most notably van der Waals
forces and higher order multipoles. However, it allows a significant
simplification of the solvation thermodynamics since all physical
properties of the solvent are expressed in terms of only two
parameters, the reduced density $\rho^*=\rho\sigma^3$ and the reduced dipole
moment $(m^*)^2 =\beta m^2/ \sigma^3$, where $m$ is the dipole moment and $\sigma$
is the diameter of the dipolar particles. Since the reduced density is
fixed to $\rho^*=0.8$ in our simulations, our results are fully defined
in terms of two parameters: the reduced cavity radius $R_0/ \sigma$ and the
polarity parameter $(m^*)^2$. The representation in terms of the
dielectric constant $\epsilon_s$ can be easily achieved as well since these
are well tabulated from our simulations as is shown in Fig.\
\ref{fig:1}. The dielectric constants were calculated from Neumann's
formalism \cite{Neumann:86} as described in detail in Ref.\
\onlinecite{DMjcp1:99}.

We will also limit our consideration to two types of electrostatic
multipoles most commonly studied in theories and applications of
solvation, point ion and point dipole
\cite{Born:20,Onsager:36,Aqvist:96}. In both cases, the corresponding
multipole is placed at the center of the spherical cavity. The
solute-solvent interaction potential is then given as $V_{0s}=q_0\phi_s$
in the case of the ion and $V_{0s}= - \mathbf{m}_0\cdot\mathbf{E}_s$ for
the dipole. In these relations, $q_0$ and $m_0$ are the charge and
dipole moment of the probe multipole and $\phi_s$ and $\mathbf{E}_s$ are,
respectively, the potential and electric field produced by the solvent
at the multipole position.

The main parameter entering the Gaussian model of solvation that we
want to monitor is the Gaussian width $\sigma^2(\beta)$. Since we want to deal
with dimensionless quantities, we will in fact calculate the
temperature reduced parameter
\begin{equation}
  \label{eq:9}
  \Gamma = \beta^2 \sigma^2(\beta) = \beta^2 \langle (\delta V_{0s})^2 \rangle_0 . 
\end{equation}
Since this parameter depends on the multipolar character of the
solute, it is convenient to take this information out and consider 
the parameter $\Delta$ such that the temperature-reduced electrostatic
energy of the solute is taken out as a multiplier
\begin{equation}
  \label{eq:10}
  \Gamma = w \Delta  .
\end{equation}
Here, the electric field of the multipole (charge or dipole) $E_0$ is
used to define the electrostatic energy
\begin{equation}
  \label{eq:11}
  w = (\beta/8\pi) \int_{\Omega} E_0(\mathbf{r})^2 d\mathbf{r} ,
\end{equation}
where the integral is taken over the solvent volume outside the
spherical cavity. 

\begin{figure}
\includegraphics*[width=7cm]{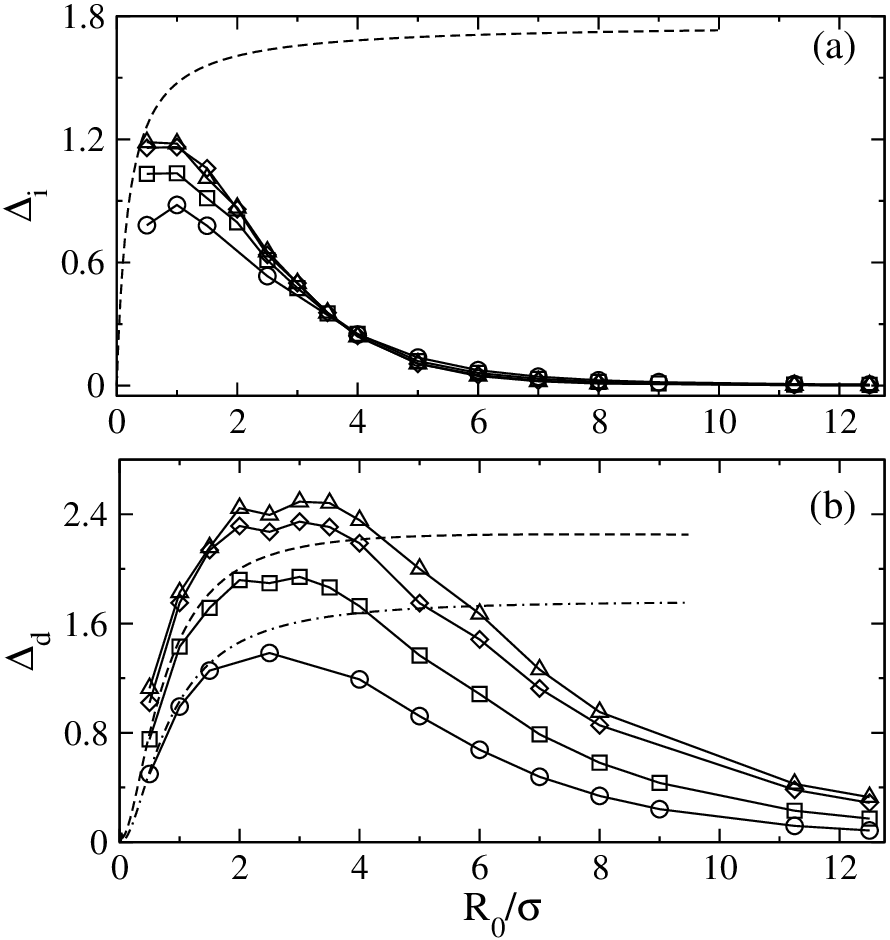}
\caption{$\Delta_i$ (a) and $\Delta_d$ (b) vs the cavity radius $R_0$ for
  $(m^*)^2=0.5$ (circles), 1.0 (squares), 2.0 (diamonds), and 3.0
  (up-triangles).  The dashed line in (a) gives the result of Eq.\
  (\ref{eq:14}) for $m^*=1.0$. The dash-dotted and dashed lines in (b)
  shows the application of Eq.\ (\ref{eq:15}) at $(m^*)^2=0.5$ and
  1.0, respectively.  }
\label{fig:2}
\end{figure}

The parameter $w$ is equal to $\beta q_0^2/(2R_0)$ for an ion and 
$\beta m_0^2/(3R_0^3)$ for a dipole, where $R_0$ is the cavity
radius. Therefore, one can calculate the parameter $\Delta$ according to 
the following relations in case on ion (subscript ``i'') or dipolar
(subscript ``d'') solutes
\begin{equation}
  \label{eq:12}
  \begin{split}
   \Delta_i & = 2 \beta R_0 \langle (\delta \phi_s)^2 \rangle_0, \\
   \Delta_d & = \beta R_0^3  \langle (\delta \mathbf{E}_s)^2 \rangle_0 . 
  \end{split}
\end{equation}
Similarly we will introduce the reduced parameter $\Delta_{ss}$ for the
components of the internal energy and entropy arising from the
alteration of the solvent-solvent interactions, $\beta \Delta\epsilon_{ss}=w\Delta_{ss}$ :
\begin{equation}
  \label{eq:17}
  \begin{split}
     \Delta_{ss}^i & = \beta^2R_0 \langle (\delta \phi_{s})^2 \delta H_0\rangle_0,\\
     \Delta_{ss}^d & = (\beta^2R_0^3/2) \langle (\delta \mathbf{E}_{s})^2 \delta H_0\rangle_0 .
  \end{split}
\end{equation}

\begin{figure}
\includegraphics*[width=7cm]{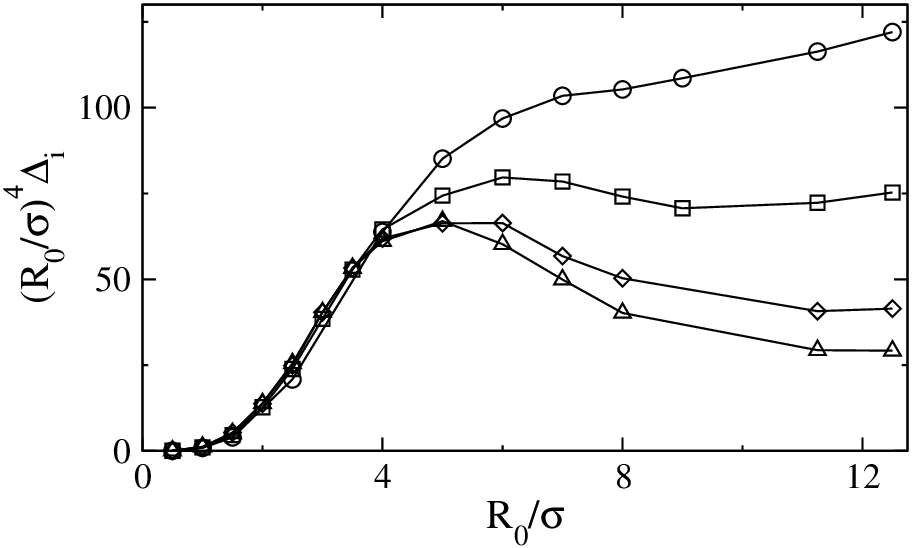}
\caption{$(R_0/ \sigma)^4\Delta_i$ vs the cavity radius $R_0$ for $(m^*)^2=0.5$
  (circles), 1.0 (squares), 2.0 (diamonds), and 3.0 (up-triangles).  }
\label{fig:3}
\end{figure}

A few analytical results from standard electrostatics \cite{Landau8}
can be used as benchmarks in calculating $\Delta_i$ and $\Delta_d$. The
continuum electrostatics of Born \cite{Born:20} and Onsager
\cite{Onsager:36} equations gives the response functions $\Delta_{i,d}$
depending only on the dielectric constant $\epsilon_s$ of the dipolar liquid:
\begin{equation}
  \label{eq:13}
   \Delta_i  = 2\left(1 - \frac{1}{\epsilon_s}\right)
\end{equation}
and
\begin{equation}
\label{eq:13-1}
   \Delta_d  = 6\frac{\epsilon_s - 1}{2\epsilon_s +1} .  
\end{equation}
In addition, several microscopic relations have been derived based on
different formulations of the liquid-state theory. A closed-form
equation for ion solvation is provided by the Ornstein-Zernike
integral equations for the ion-dipole mixture solved in the
mean-spherical approximation (MSA) \cite{Chan:79}:
\begin{equation}
  \label{eq:14}
  \Delta_i = \frac{2R_0}{R_0 + \Lambda_L}\,\left(1- \frac{1}{\epsilon_s}\right) .
\end{equation}
In this equation, $\Lambda_L=3\sigma \xi/(1+4\xi)$ is the correlation length of
longitudinal polarization fluctuations of a dipolar liquid and $\xi$ is the
MSA polarity parameter \cite{Wertheim:71}. 

An analogous MSA solution exists for the mixture of dipolar particles
of different size \cite{Freasier:79} which gives the parameter $\Delta_d$.
Truncated perturbation expansions \cite{Larsen:77} are however known
to work better in this case with the result \cite{DMjcp1:99,DMjpca:04}
\begin{equation}
  \label{eq:15}
  \Delta_d = 6\left(\frac{R_0}{R_{\text{eff}}} \right)^3
  \frac{y}{1+\kappa(y,r_{0s})y\sigma^3 I_{0s}^{(3)}/R_{\text{eff}}^3 }.
\end{equation}
Here, $r_{0s}=R_0/ \sigma + 0.5$ is the reduced distance of the closest
approach of the liquid molecules to the cavity and $y=(4\pi/9)\beta m^2 \rho$
is the standard density of dipoles in the dipolar liquid
\cite{Boettcher:73}, $\rho$ is the liquid number density. In addition,
$I_{0s}^{(3)}(r_{0s},\rho^*)$ is the three-particle perturbation integral
which is a function of the liquid density and $r_{0s}$ and
$R_{\text{eff}}(r_{0s},\rho^*)$ is the effective radius of the cavity
\begin{equation}
  \label{eq:16}
  R_{\text{eff}}^{-3}(r_{0s},\rho^*) = 3 \int_{0}^{\infty} \frac{dr}{r^4}
  g_{0s}^{(0)}(r) . 
\end{equation}
In this equation $g_{0s}^{(0)}(r)$ is the hard-sphere distribution
function of the liquid particles as a function of the distance $r$ to
the cavity center. All functions $R_{\text{eff}}(r_{0s},\rho^*)$,
$I^{(3)}(r_{0s},\rho^*)$, and $\kappa(y,r_{0s})$ are given as analytical
functions of the corresponding parameters in Ref.\
\onlinecite{DMjpca:04}.

\section{Results}
Our simulations have produced an unexpected result. We found that the
scalings of electrostatic fluctuations and the corresponding
chemical potentials with the cavity size do not follow the predictions
of both the continuum electrostatics and microscopic solvation models
in case of large cavities. The results are shown in Fig.\
\ref{fig:2}. As is seen, the parameter $\Delta_i$ decays much faster than
the expected $1/R_0$ scaling for all cavities greater than the size of
the solvent particle. The large cavity scaling does not follow any
universal law, but instead depends on the polarity (parameter $m^*$)
of the liquid (Fig.\ \ref{fig:3}). For the liquid polarities studied
here, the large-cavity scaling of $\Delta_i$ is approximately
$1/R_0^{4-6}$.  Fluctuations of the electric field at the cavity
center, representing dipole solvation, do not deviate that
dramatically from the traditional expectations, but the parameter
$\Delta_d$ still decays to zero instead of leveling off as suggested by
Eqs.\ (\ref{eq:13}) and (\ref{eq:15}). In fact, $\Delta_d$ follows Eq.\
(\ref{eq:15}) quite well up to the cavity size about 4--5 times larger
than the liquid particle, but then starts to drop following
qualitatively the trend seen for the potential fluctuations. Continuum
electrostatics [Eq.\ (\ref{eq:13})] fails both qualitatively and
quantitatively for electrostatic fluctuations of both the potential
and the electric field.

There is a slight dependence of the variances on the number of
particles in the simulation box. The variances extrapolated to $N\to \infty$
from simulations done at various system sizes are listed in Table
\ref{tab:1} in the Appendix. This dependence does not affect any
qualitative conclusions we make here. Since extrapolation to $N\to \infty$
creates a scatter of points, the results presented in Fig.\
\ref{fig:2} refer to a given system size only.

\begin{figure}
  \centering
  \includegraphics*[width=7cm]{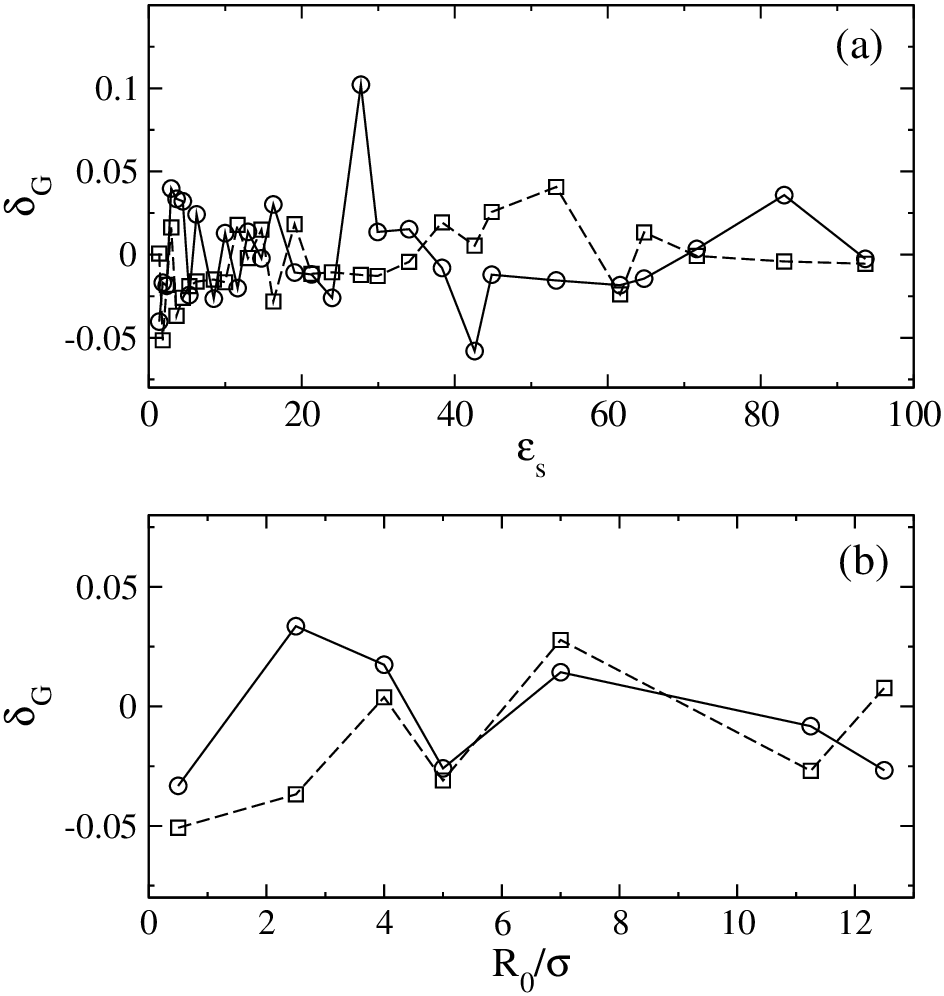}
  \caption{Non-gaussianity parameter $\delta_G$ [Eq.\ (\ref{eq:19})]
    vs $\epsilon_s$ (a) and $R_0$ (b). Points represent probe ions
    (circles) and probe dipoles (squares) for $R_0/ \sigma = 2.5$ (a) and 
    $(m^*)^2=0.5$ (b).  }
  \label{fig:4}
\end{figure}

With the dramatic failure of some very basic expectations regarding
electrostatic fluctuations, as is shown in Fig.\ \ref{fig:2}, one
wonders if the Gaussian approximation for the distribution of the
electrostatic interaction energies fails for large cavities. We have
tested this question by looking at the non-gaussianity parameter for
both potential and field fluctuations:
\begin{equation}
  \label{eq:19}
  \delta_G = \frac{\langle (\delta V_{0s})^4\rangle_0 }{\langle(\delta V_{0s})^2\rangle_0^2} - 3.  
\end{equation}
This parameter was found to be around zero, as expected for the
Gaussian noise, within about 5\% of the simulation uncertainties (Fig.\
\ref{fig:4}). The Gaussian approximation therefore seems reliable for
our parameters database.

\begin{figure}[ht]
\includegraphics*[width=7cm]{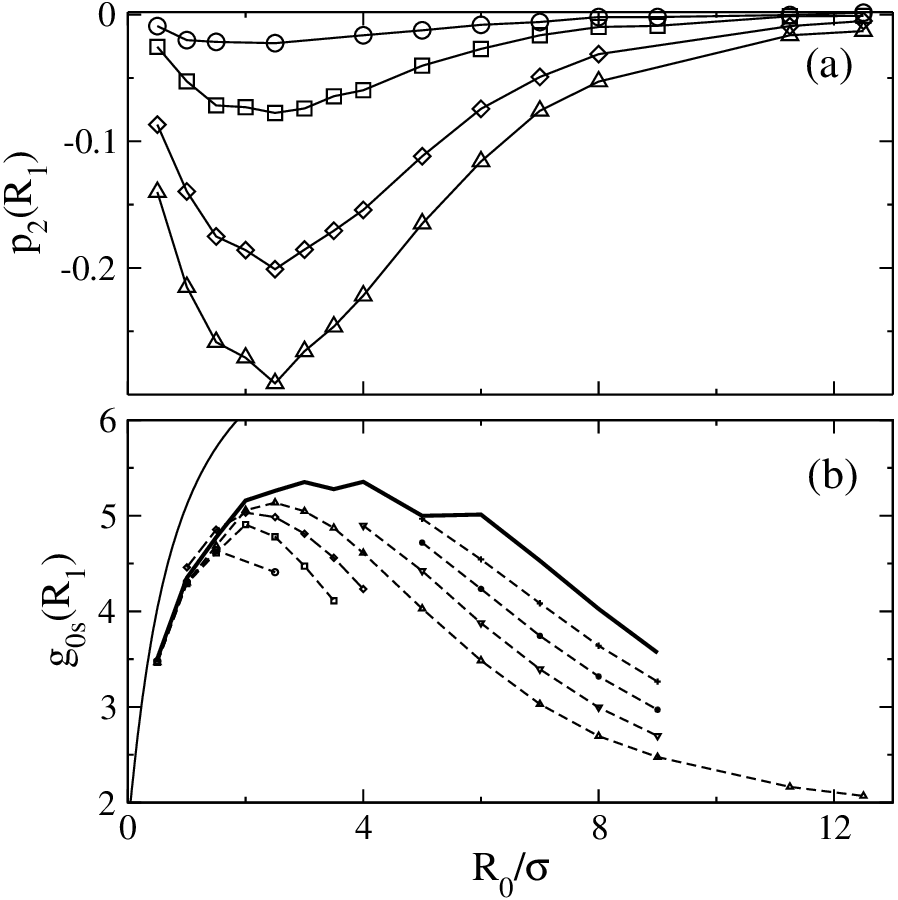}
\caption{(a): the orientational order parameter vs the cavity
  size for different polarities of the solvent, $(m^*)^2=0.5$
  (circles), 1.0 (squares), 2.0 (diamonds), and 3.0 (up-triangles).
  (b): contact value of the radial distribution function at
  $R_1 = R_0 + \sigma/2$ vs the cavity radius.  Shown are the results for
  different number of particles in the simulation box $N=256$
  (circles), 500 (squares), 864 (diamonds), 1372 (up-triangles), 2048
  (down-triangles), 2916 (stars), 4000 (pluses).  Extrapolation to $N\to
  \infty$ is shown by bold solid line. The dashed lines connect the points.
  The thin solid line gives the contact value of the distribution
  function in the hard-spheres mixture from Ref.\
  \onlinecite{DMjcp3:97}. }
\label{fig:5}
\end{figure}

In order to gain more insight into the origin of our observations, we
have calculated two local parameters related to the orientational and
density structure of the liquid/cavity interface. Figure \ref{fig:5}a
shows the second-rank orientational order parameter of the permanent
dipoles in the first solvation shell at the cavity surface:
\begin{equation}
  \label{eq:20}
  p_2(r) = \left\langle  \sum_j P_2 (\mathbf{\hat r}_j \cdot \mathbf{\hat e}_j
    ) \delta(\mathbf{r}_j - \mathbf{r}) \right\rangle .              
\end{equation}
Here, $P_2(x)$ is the second Legendre polynomial, $\mathbf{\hat
  r}_j=\mathbf{r}_j/r_j$ is the unit vector in the direction of the
liquid particle $j$, and $\mathbf{\hat e}_j$ is the unit vector along
its dipole moment.  The orientational order parameter shown in Fig.\
\ref{fig:5}a is calculated by limiting the distance $r$ to liquid
particles residing in the cavity's first solvation shell where it
indicates the existence of a preferential orientational order.  The
first-rank orientational parameter, based on the first-order Legendre
polynomial, is identically zero thus implying that there is no net
dipolar polarization at the cavity surface. This result is distinct
from the water surface where water's large quadrupole moment is
responsible for asymmetry \cite{Sokhan:97}.

\begin{figure}
  \centering
  \includegraphics*[width=6cm]{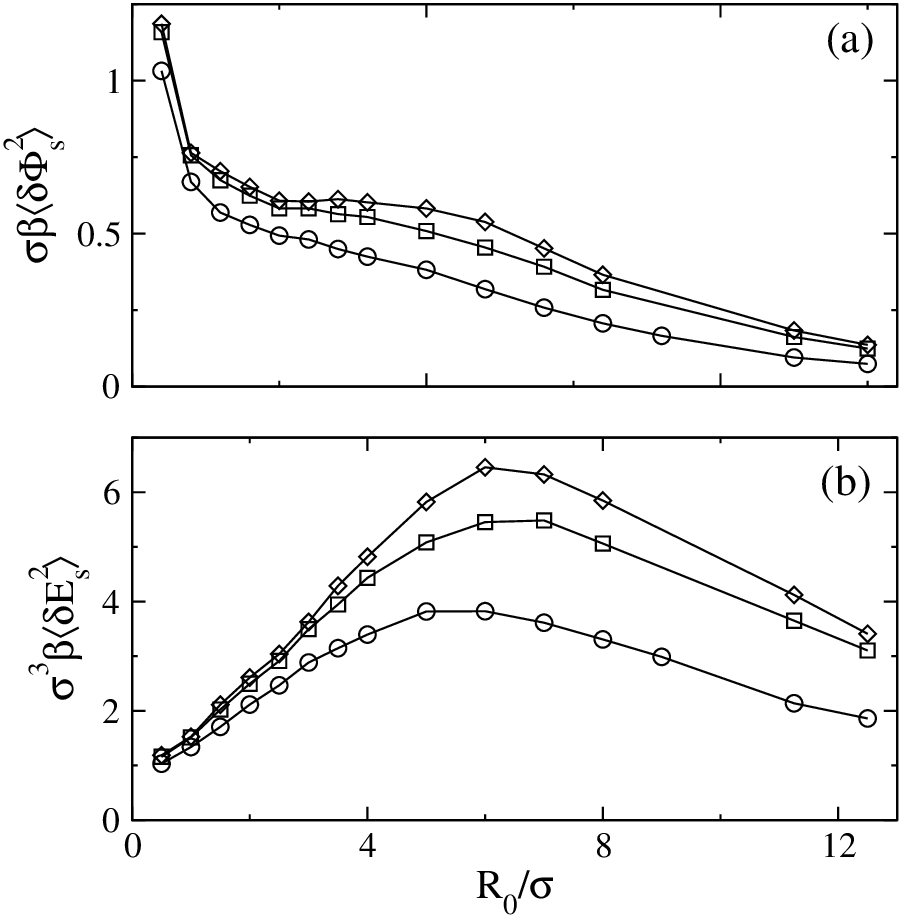}
  \caption{$\sigma \beta \langle(\delta \phi_s)^2\rangle$ (a) and $\sigma^3 \beta \langle (\delta \mathbf{E}_s)^2\rangle$ (b)
    vs the cavity size for probe charge and dipole located the
    distance $\sigma/2$ from the cavity surface. The points refer to
    $(m^*)^2=1.0$ (circles), 2.0 (squares), and 3.0 (diamonds);
    $N=1372$.  }
  \label{fig:6}
\end{figure}

As the cavity gets larger the solvent dipoles find it more
energetically favorable to orient parallel to the interface, as was
also observed for 2D dipolar liquids \cite{Weis:02}, for water at
cavity surfaces \cite{Rajamani:04} and liquid-vapor interfaces
\cite{Sokhan:97}, and for interfaces of dipolar liquids \cite{Frodl:92}
from density-functional calculations. However, this preferential
orientational order starts to dissolve with a further increase of the
cavity size, after gaining maximum for the cavity about five times
larger than the solvent particle. This decay is related to the onset
of softening of the first solvation shell indicated by the contact
value of the pair cavity-solvent distribution function shown in Fig.\
\ref{fig:5}b.

The contact value of the pair distribution function first rises as
expected for a hard-sphere impurity in densely packed hard spheres
\cite{DMjcp3:97} (solid line in Fig.\ \ref{fig:5}b), but then starts
to drop. This drop appears at approximately the same value $R_0/ \sigma \simeq
2-2.5$ as both the downward turn of the orientational order parameter
and the onset of deviation of the electric field fluctuations from the
traditional predictions (Fig.\ \ref{fig:2}). We therefore can conclude
that the observed change in the character of the electrostatic
fluctuations is related to softening of the liquid/cavity interface,
which also loosens the energetic push for a specific dipolar order.
We note, however, that the peak of the distribution function stays at
the closest-approach value $R_1 = R_0 + \sigma/2$ and thus no dewetting
\cite{HummerPRL:98} of the cavity interface occurs.

That the decay of the solvation energies is related to the softening
of the interface is also seen from probing the fluctuations of the
potential and field close to the cavity interface. Figure \ref{fig:6}
shows the corresponding quantities for a point within the cavity kept
one solvent radius $\sigma/2$ away from the interface once the cavity size
is increased. Again, simple electrostatic arguments suggest that the
solvation energetics should approach that for a probe charge or dipole
next to an infinite dielectric wall. Depending on how the dielectric
interface is defined, by the cavity boundary or by the distance of the
closest approach, continuum electrostatics predicts \cite{Landau8} for
$\sigma\beta\langle (\delta \phi_s)^2\rangle $ the value between $(\epsilon_s -1)/(\epsilon_s +1)$ and $0.5(\epsilon_s
-1)/(\epsilon_s + 1)$.  The observed dependence does seem to inflect into a
plateau at the level consistent with this prediction at intermediate
cavity size, but then starts to decay.  This decay is however much
more gentle than in Fig.\ \ref{fig:2} indicating that the area next to
the interface is effectively stronger solvating than the part of the
hollow space closer to the cavity center.

\begin{figure}
  \centering
  \includegraphics*[width=6cm]{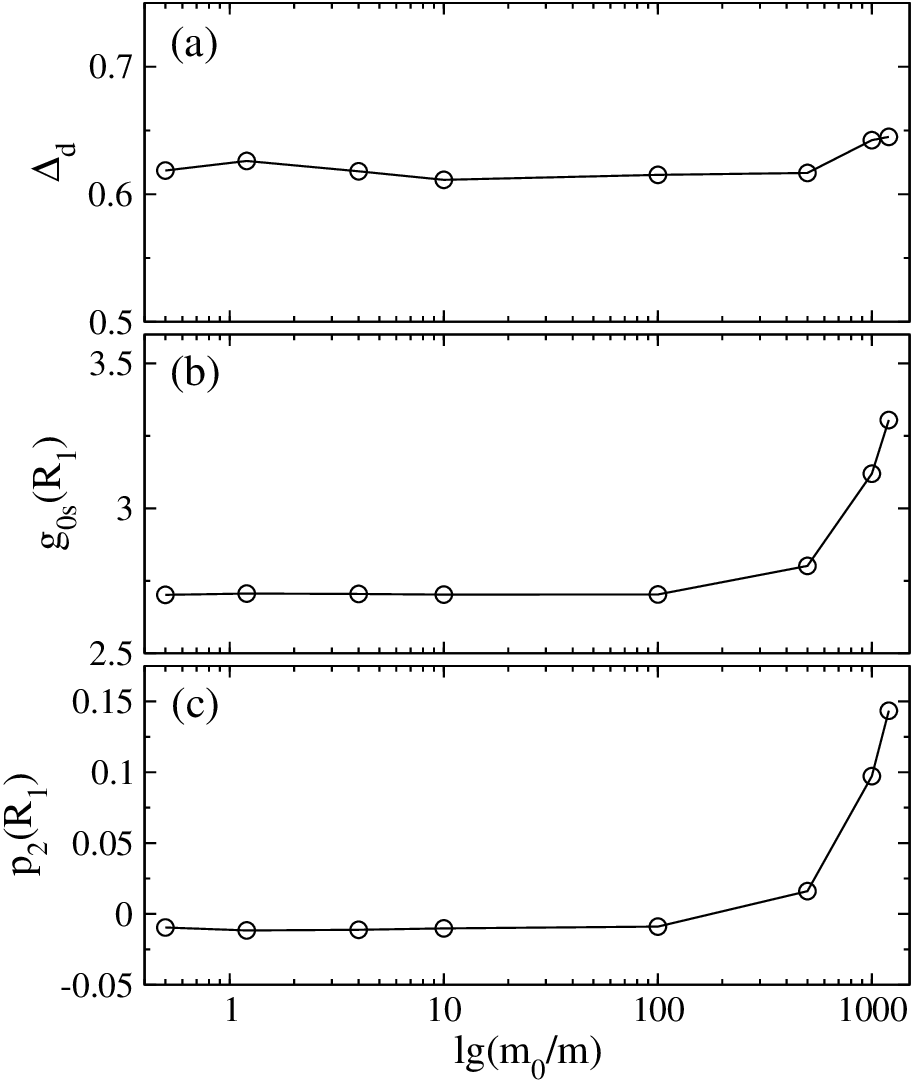}
  \caption{Response function $\Delta_d$ (a), contact value of the
    cavity-solvent pair distribution function $g_{0s}(R1)$ (b), and
    the orientational order parameter of the dipoles in the first
    solvation shell $p_2(R_1)$ (c) vs the magnitude of the solute
    dipole at the cavity center $m_0$. Points are the results of MC
    simulations with $R_0/ \sigma=9.0$, $(m^*)^2=1.0$, and $N=2048$. The
    lines connect the simulation points.  }
  \label{fig:7}
\end{figure}

The Gaussian approximation is a central part of our thermodynamic
arguments and so we have done an additional test of its consistency
also offering some deeper insights into the nature of electrostatic
response functions. Since the chemical potential of solvation is given
by the variance of the solute-solvent interaction potential [Eq.\
(\ref{eq:6})], it becomes quadratic in a test multipole used to probe
the electrostatic fluctuations.  This result, known as the linear
response approximation \cite{Aqvist:96}, suggests that the response
function, obtained as the second derivative of $\mu_{0s}$ in the
corresponding multipole, does not depend any more on the magnitude of
that multipole. It also implies that $\Delta_{i,d}$ can be obtained from
simulations of empty cavities but also from simulations involving
actual multipoles inside the cavity.  The chemical potential of
solvation and corresponding parameters $\Delta_{i,d}$ are then calculated
from the average solute-solvent interaction energy using Eq.\
(\ref{eq:18}). Since such simulations involving the probe charge are
not straightforward due to the breakdown of the system neutrality and
the related difficulty of using the Ewald sums
\cite{Leeuw:80,Hummer:96}, we have done simulations of point dipoles
of varying magnitude placed at the cavity's center. The results are
shown in Fig.\ \ref{fig:7} for the cavity size above the threshold
seen in Fig.\ \ref{fig:2}, $R_0/ \sigma= 9.0$, and the simulation box
containing $N=2048$ solvent particles.  There we show the parameter
$\Delta_d$ calculated from $\langle V_{0s}\rangle$ at the varying magnitude of the
solute dipole $m_0$. Figures \ref{fig:7}b and \ref{fig:7}c also
present the corresponding contact values of the cavity-solvent pair
distribution function $g_{0s}(R_1)$ and the orientational order
parameter $p_2(R_1)$. The response function $\Delta_d$ stays constant
almost in the entire range of $m_0$ studied, starting to rise when the
dipole inside the cavity exceeds the solvent dipole by three orders of
magnitude. This rise is a reflection of the change in the microscopic
structure of the interface as the first solvation shell gets
stiffer under the pull of the solute dipolar field and the
first-shell dipoles start to reorient along the field of the solute
dipole. The observed changes in the functions $g_{0s}(R_1)$ and
$p_2(R_1)$ are, however, much greater than the corresponding change in
$\Delta_d$ testifying to the collective nature of the solvent dipolar
response effectively depressing changes in the microscopic structure
of the first solvation shell.

\begin{figure}
  \centering
  \includegraphics*[width=6cm]{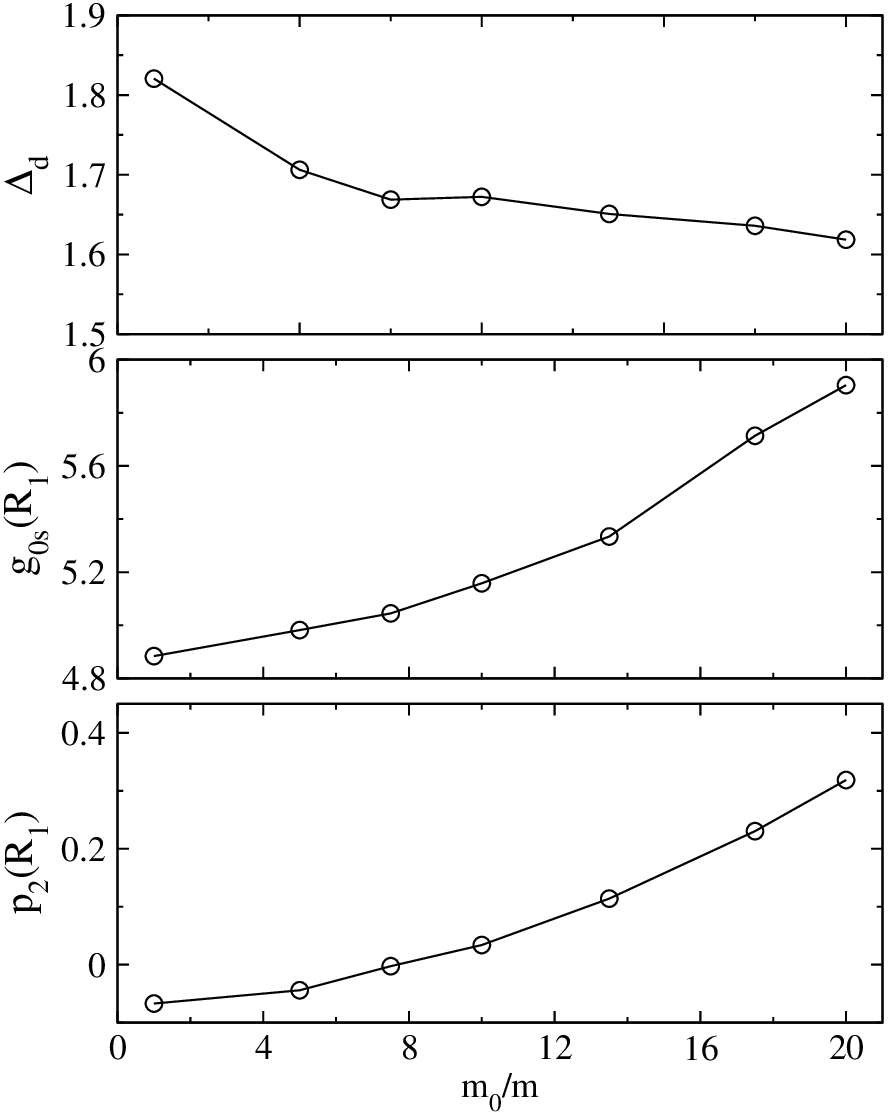}
  \caption{Same as in Fig.\ \ref{fig:7} for $R_0/ \sigma=1.5$. }
  \label{fig:8}
\end{figure}

In Fig.\ \ref{fig:8} we show the same data as in Fig.\ \ref{fig:7},
but obtained at a much smaller cavity size $R_0/ \sigma= 1.5$. Here, the
change in the local structure with increasing the solute dipole is
more pronounced and $\Delta_d$ starts to show a dependence on the
magnitude of the probe dipole signalling the appearance of nonlinear
solvation effects. The variation in the response function is still
mostly within 10\% and can be accounted for by nonlinear extensions of
dipolar solvation models \cite{DMjcp1:99}. We need to stress, however,
that the Gaussian approximation appears to be robust for large cavities
which are of main interest for us here.

\begin{figure}
  \centering
  \includegraphics*[width=6cm]{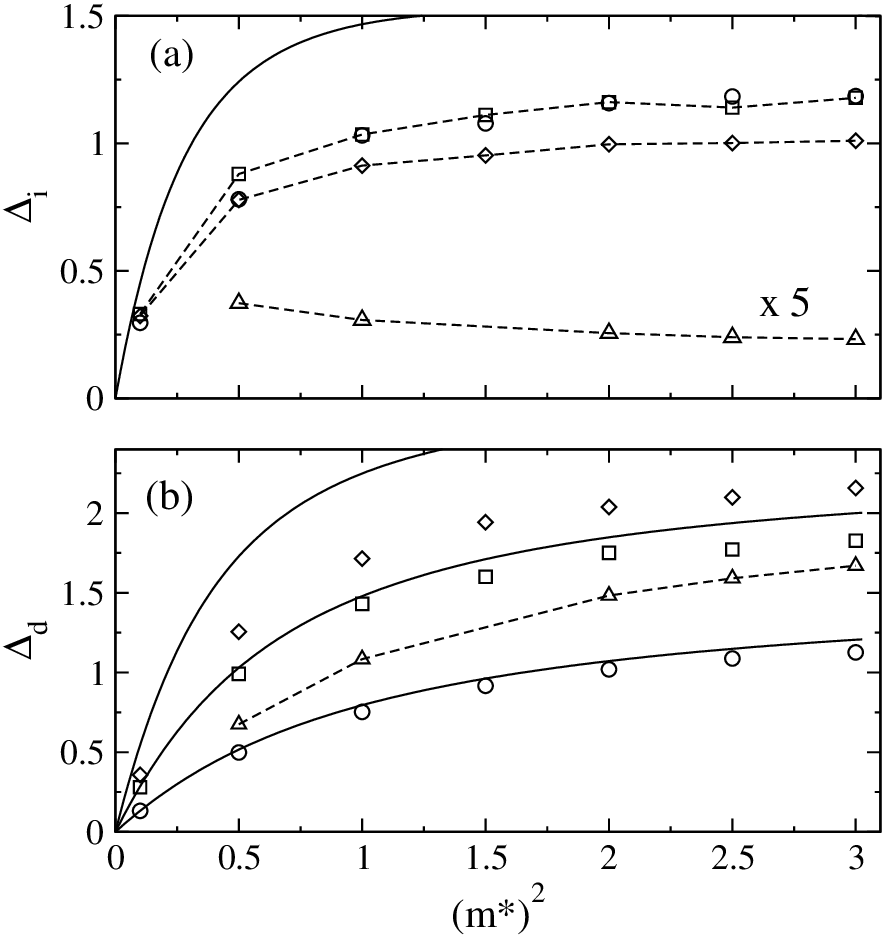}
  \caption{ $\Delta_i$ (a) and $\Delta_d$ (b) as functions of $(m^*)^2$ for
    $R_0/\sigma = 0.5$ (circles), 1.0 (squares), 1.5 (diamonds), and 6.0
    (up-triangles). The solid line in (a) shows the result of using
    Eq.\ (\ref{eq:14}) at $R_0/ \sigma=1.0$. The solid lines in (b) show
    the result of Eq.\ (\ref{eq:15}) for $R_0/ \sigma = 0.5$, 1.0, and 6.0
    (from down up); the dashed lines in (a) and (b) connect the
    points.  The data for $\Delta_i$ at $R_0/\sigma = 6.0$ (up-triangles)
    in (a) have been multiplied by a factor of 5 to bring them to the
    scale of the plot. The simulation points were obtained at $N=1372$
    dipolar hard spheres in the box. }
  \label{fig:9}
\end{figure}

We next turn to the dependence of the cavity response functions on the
liquid polarity.  Figure \ref{fig:9}a shows the dependence of $\Delta_{i}$
on the solvent dipole moment. For a small cavity size, when the
standard scaling with the cavity size is expected to apply, the
dependence of $\Delta_i$ on polarity does not show a saturation predicted
by continuum electrostatics [Eq.\ (\ref{eq:13})]. This saturation
appears for a slightly larger cavity, but, as seen for a still larger
cavity, it is simply en route to become a decreasing function of
polarity for the largest cavities studied here.  We can therefore
conclude that there is no range of parameters where both the size
scaling and the dependence on polarity predicted by the continuum
electrostatics for the potential fluctuations are satisfied even at
the qualitative level, not to mention the fact that the predicted
values are significantly off.

The saturation predicted by the Onsager equation for dipole solvation
[Eq.\ (\ref{eq:13-1})] is never reached. In contrast to the potential
fluctuations, the variance of the field is a uniformly increasing
function with increasing solvent dipole for all cavity sizes studied
here. A similar trend, for a narrower range of parameters, was
previously observed by us \cite{DMjpca:02} and it manifests itself in
the solvation dynamics uniformly slower than continuum predictions
\cite{DMjcp1:05}. The results for $\Delta_d$ from Eq.\ (\ref{eq:15}) are
shown by the solid lines in Fig.\ \ref{fig:9}b. As expected, there is a
good agreement with the simulations for small cavities, but then the
theory fails when the regime of solvation changes and $\Delta_d$ turns
downward as in Fig.\ \ref{fig:2}.

\begin{figure}
  \centering
  \includegraphics*[width=6cm]{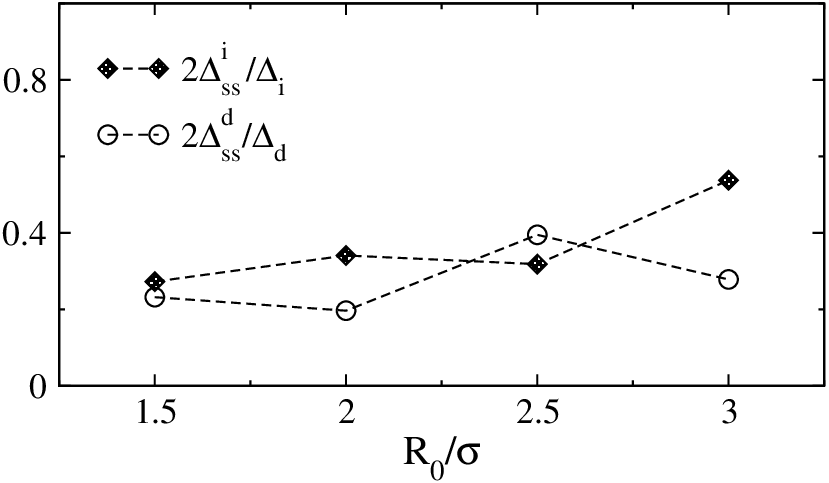} 
  \caption{The ratio of the solute-solvent and solvent-solvent
    components of the solvation entropy vs the cavity radius
    calculated for charge and dipole probe multipoles. }
  \label{fig:10}
\end{figure}

In Fig.\ \ref{fig:10} we show the results of calculations of the
solvent-solvent component of the solvation entropy [Eq.\
(\ref{eq:7})]. Within the Gaussian approximation, the ratio of the
solvent-solvent component of the solvation entropy, $Ts_{ss}=\Delta e_{ss}$,
and the solute-solvent component, $Ts_{0s}= - \beta \sigma^2(\beta)/2$, is given as
the ratio of the corresponding reduced response functions 
\begin{equation}
  \label{eq:22}
  \chi_s = -\frac{s_{ss}^{i,d}}{s_{0s}^{i,d}}=
  \frac{2\Delta_{ss}^{i,d}}{\Delta_{i,d}} .
\end{equation}
As is seen, for both the ionic and dipole solvation, there is a
compensation between the ordering of the solvent by the solute,
expressed by always negative $s_{0s}$, and the disordering of the
solvent structure, expressed by positive $s_{ss}$. This compensation
is however far from complete, in contrast to a much stronger
compensation found for aqueous solvation \cite{DMjpca1:06}. The
overall entropy of electrostatic solvation is therefore
negative. Since the parameter $\chi_s$ in Eq.\ (\ref{eq:22}) depends
weakly on the cavity size, the dramatic change in the character of
solvation found here for $\sigma^2(\beta)$ will be reflected in both the
enthalpy and entropy of electrostatic solvation which are often more
accessible experimentally than solvation free energies. Very little is
currently known about the magnitude of $\chi_s$ \cite{Ben-AmotzAcc:08},
in particular for large solutes. Our recent MD simulations of the
redox entropy of metalloprotein plastocyanin \cite{DMjcp2:08} have
produced $\chi_s \simeq 0.4$ ($R_0/ \sigma \simeq 5.8$), although it is not clear if the
Gaussian approximation is applicable to the protein electrostatics.

\section{Discussion}
In this paper we have suggested to study polar solvation by using Eq.\
(\ref{eq:3}) which states that all the information required to
calculate the solvation thermodynamics is contained in the
distribution of electrostatic interaction energies around a fictitious
solute with the solute-solvent electrostatic coupling switched
off. This equation is exact and the approximation adopted here is that
the distribution function $P(\epsilon)$ can be approximated by a
Gaussian. The distribution $P(\epsilon)$ can generally be written as
\begin{equation}
  \label{eq:21}
  P(\epsilon) \propto \exp\left[\beta \omega(\epsilon)  \right]
\end{equation}
and then the integral over $\epsilon$ in Eq.\ (\ref{eq:3}) can be taken by
the steepest descent around the stationary point $\epsilon_0$ defined by the
condition $\omega'(\epsilon_0) = 1$. The Gaussian approximation is then equivalent
to assuming all the terms except the linear one can be dropped from
the series expansion of $\omega'(\epsilon)$ in powers of $(\epsilon-\epsilon_0)$.

Our simulations have not identified any significant deviations from
non-gaussianity.  Extensive simulations done with ionic and dipolar
solutes over the last decades
\cite{Aqvist:96,Kuharski:88,DMjpca:02,Blumberger:05} have also
resulted in the conclusion that the Gaussian picture is an accurate
one implying that $P(\epsilon)$ is globally a Gaussian function. However, one
can argue that we could not sample sufficiently around $\epsilon_0$ and thus
cannot assess the deviations from Gaussianity.  While that might be
true for strong solute-solvent interactions, for which a significant
data-base pointing otherwise exists \cite{DMjpca:02}, energy $\epsilon_0$ is
expected to decrease with increasing the cavity size and the Gaussian
approximation is expected to become increasingly accurate (as indeed
seen from comparing Figs.\ \ref{fig:7} and \ref{fig:8}).  However, it
is in this range of large cavities, almost completely neglected in
previous studies of electrostatic solvation, that we found the most
dramatic deviations from the traditional expectations.
 
The main funding of this study is that electrostatic solvation by
polar liquids changes its regime at the size of the cavity about 4-5
times larger than the size of the solvent particle. The regime of
small cavities can be reasonably understood with molecular solvation
models and in particular the results for the electric field
fluctuations (probe dipole) are in a very good quantitative agreement
with the results of perturbation solvation models. The regime of large
cavities is dramatically different and cannot be described by the
models traditionally employed for solvation problems.

What we have observed here is a dramatic decay of the solvation
strength in the middle of the cavity, much faster than expected from
both the continuum electrostatics and microscopic solvation
models. For instance, the variance of the electrostatic potential
decays as $1/R_0^{4-6}$ instead of the expected $1/R_0$ scaling. The
core of a growing hollow cavity thus becomes non-polar much faster
than previously anticipated. What it practically means is that there
is very little solvation stabilization for charges inside a large
mesoscale object. This might be a reason why natural systems requiring
hydration of large molecular assemblies (proteins, etc.)  rely on
solvation of surface charges for which much slower decay of solvating
power due to softening of the interface was found here.  In
application to the problem of protein folding, this observation
implies a very strong driving force for placing ionized residues and
cofactors stabilizing protein solvation closer to the interface.

\acknowledgments This research was supported by the National Science
Foundation (CHE-0616646). CPU time for parallel MC simulations was provided
by ASU's Center for High Performance Computing. 

\appendix
\section{Simulation protocol and results}
MC simulations were performed with the standard NVT Metropolis
algorithm.  The initial configuration was constructed starting from an
\textit{fcc} lattice of liquid hard-spheres of diameter $\sigma$ and
density $\rho^*=0.8$.  The hard-sphere solute/cavity was then ``grown''
in the center of the simulation box by increasing the initial cavity
diameter of $0.5\sigma$ with $0.002\sigma$ increment, adjusting $\sigma$ to ensure
constant density, and moving the solvent particles according to the
Metropolis algorithm. After the solute/cavity was constructed, the
initial configuration was created from $10^5-10^6$ parallel steps
(using OpenMPI) producing different initial configurations for each
processor. The subsequent runs were then carried out on each processor
separately thus minimizing interprocessor communications. To guarantee
the Markovian statistics, the random number generators used in the MC
moves were seeded independently between the processors. This
implementation has resulted in a linear scaling of the program output
with the number of processors. The production runs of $(1-5)\times10^6$
steps were performed on 10 processors per $(m*)^2$ per cavity size.

The simulation protocol employed the minimum image convention and the
reaction-field correction \cite{Allen:96} for the cut-off of dipolar
interactions at one-half of the cubic simulation box. Ewald sums
\cite{Leeuw:80} were also tested and gave results identical within
simulation uncertainties. The reaction-field correction was preferred
due to better performance.  The dependence on the simulation box size
was carefully checked in particular since growing cavity required
larger number of liquid particles to eliminate finite-size
effects. The number of particles $N$ was varied in the range $N=108$,
256, 500, 864, 1372, 2048, 2916, and 4000 depending on the cavity
size. The representative results for $\Delta_i$ and $\Delta_d$ listed in Table
\ref{tab:1} were obtained by averaging over several simulation runs
with different box sizes and also by extrapolating the plots of
corresponding values vs $1/N$ to the $N\to \infty$ limit.

\begingroup 

\squeezetable

\begin{table*}
\begin{center}
\caption{Values of $\Delta_i$ and $\Delta_d$ for $(m^*)^2=0.5, 1.0, 2.0, 3.0$,
  $\epsilon=3.63, 8.51, 29.9, 93.7$, respectively.  Extrapolations (ext.)
  were done with $N\!=\!108, 256, 500,  864, 1372, 2048,  2916, 4000$
  data  when available, linearly fitting  $\Delta_{i,d}$ vs.\ $1/N$ and taking the
  intercept.  The system sizes used for the extrapolations are given
  in the footnotes.  \label{tab:1}}
\begin{tabular}{cccccccccc}
\hline\hline
 & & \multicolumn{8}{c}{$(m^*)^2$} \\
\cline{3-10}
 & & \multicolumn{2}{c}{0.5} & \multicolumn{2}{c}{1.0} &
 \multicolumn{2}{c}{2.0} & \multicolumn{2}{c}{3.0} \\
$R_0/\sigma$ & $N$ & $\Delta_i$ & $\Delta_d$ & $\Delta_i$ & $\Delta_d$ &
$\Delta_i$ & $\Delta_d$ & $\Delta_i$ & $\Delta_d$ \\
\hline
0.5 & 1372 &~ 0.781 ~&~ 0.498 ~&~ 1.031 ~&~ 0.753 ~&~
    1.158 ~&~ 1.020 ~&~ 1.185 ~&~ 1.126 ~\\
    & ext.\footnotemark[1] &~ 0.814 ~&~ 0.572 ~&~ 1.052 ~&~ 0.761 ~&~
    1.210 ~&~ 1.030 ~&~ 1.265 ~&~ 1.153 ~\\ 
1.0 & 1372 &~ 0.892 ~&~ 1.001 ~&~ 1.035 ~&~ 1.431 ~&~ 1.184 ~&~ 1.801
    ~&~ 1.240 ~&~ 1.939 ~\\
    & ext.\footnotemark[2] &~ 0.863 ~&~ 1.009 ~&~ 1.050 ~&~ 1.449 ~&~
    1.171 ~&~ 1.768 ~&~ 1.198 ~&~ 1.854 ~\\
1.5 & 1372 &~ 0.778 ~&~ 1.256 ~&~ 0.936 ~&~ 1.760 ~&~ 1.058 ~&~ 2.137
    ~&~ 1.073 ~&~ 2.286 ~\\
    & ext.\footnotemark[3] &~ 0.834 ~&~ 1.292 ~&~ 1.011 ~&~ 1.772 ~&~
    1.084 ~&~ 2.112 ~&~ 1.103 ~&~ 2.238 ~\\
2.0 & 1372 &~ -- ~&~ -- ~&~ 0.795 ~&~ 1.918 ~&~ 0.859 ~&~ 2.314 
    ~&~ 0.866 ~&~ 2.444 ~\\
    & ext.\footnotemark[3] &~ -- ~&~ -- ~&~ 0.862 ~&~ 1.965 ~&~ 0.844
    ~&~ 2.300 ~&~ 0.859 ~&~ 2.394 ~\\
2.5 & 1372 &~ 0.533 ~&~ 1.384 ~&~ 0.632 ~&~ 1.955 ~&~ 0.670 ~&~ 2.365 
    ~&~ 0.666 ~&~ 2.513 ~\\
    & ext.\footnotemark[3] &~ 0.619 ~&~ 1.491 ~&~ 0.692 ~&~ 2.079 ~&~
    0.695 ~&~ 2.470 ~&~ 0.710 ~&~ 2.627 ~\\
3.0 & 1372 &~ -- ~&~ -- ~&~ 0.475 ~&~ 1.941 ~&~ 0.499 ~&~ 2.345
    ~&~ 0.499 ~&~ 2.492 ~\\
    & ext.\footnotemark[4] &~ -- ~&~ -- ~&~ 0.612 ~&~ 2.122 ~&~ 0.550
    ~&~ 2.468 ~&~ 0.556 ~&~ 2.587 ~\\
3.5 & 1372 &~ -- ~&~ -- ~&~ 0.352 ~&~ 1.864 ~&~ 0.355 ~&~ 2.307
    ~&~ 0.355 ~&~ 2.482 ~\\
    & ext.\footnotemark[4] &~ -- ~&~ -- ~&~ 0.459 ~&~ 2.129 ~&~ 0.376
    ~&~ 2.539 ~&~ 0.375 ~&~ 2.578 ~\\
4.0 & 1372 &~ 0.249 ~&~ 1.191 ~&~ 0.252 ~&~ 1.723 ~&~ 0.242 ~&~ 2.188
    ~&~ 0.239 ~&~ 2.356 ~\\
    & ext.\footnotemark[5] &~ -- ~&~ -- ~&~ 0.495 ~&~ 2.186 ~&~ 0.528
    ~&~ 2.597 ~&~ 0.533 ~&~ 2.699 ~\\
5.0 & 1372 &~ 0.136 ~&~ 0.927 ~&~ 0.124 ~&~ 1.431 ~&~ 0.112 ~&~ 1.830 
    ~&~ 0.107 ~&~ 2.000 ~\\
    & ext.\footnotemark[6] &~ 0.299 ~&~ 1.393 ~&~ 0.310 ~&~ 2.004 ~&~
    0.251 ~&~ 2.388 ~&~ 0.246 ~&~ 2.556 ~\\
6.0 & 1372 &~ 0.075 ~&~ 0.675 ~&~ 0.061 ~&~ 1.084 ~&~ 0.051 ~&~ 1.483 
    ~&~ 0.046 ~&~ 1.670 ~\\
    & ext.\footnotemark[7] &~ 0.298 ~&~ 1.449 ~&~ 0.296 ~&~ 2.061 ~&~
    0.236 ~&~ 2.484 ~&~ 0.233 ~&~ 2.553 ~\\
7.0 & 1372 &~ 0.043 ~&~ 0.478 ~&~ 0.033 ~&~ 0.789 ~&~ 0.024 ~&~ 1.124
    ~&~ 0.021 ~&~ 1.263 ~\\
    & ext.\footnotemark[7] &~ 0.192 ~&~ 1.233 ~&~ 0.183 ~&~ 1.834 ~&~
    0.133 ~&~ 2.203 ~&~ 0.125 ~&~ 2.373 ~\\
8.0 & 1372 &~ 0.026 ~&~ 0.338 ~&~ 0.018 ~&~ 0.581 ~&~ 0.012 ~&~ 0.855 
    ~&~ 0.010 ~&~ 0.952 ~\\
    & ext.\footnotemark[7] &~ 0.128 ~&~ 1.003 ~&~ 0.114 ~&~ 1.540 ~&~
    0.054 ~&~ 1.790 ~&~ 0.048 ~&~ 2.004 ~\\
9.0 & 1372 &~ 0.017 ~&~ 0.241 ~&~ 0.011 ~&~ 0.433 ~&~ -- ~&~ --
    ~&~ -- ~&~ -- ~\\
    & ext.\footnotemark[7] &~ 0.087 ~&~ 0.803 ~&~ 0.071 ~&~ 1.258 ~&~
    -- ~&~ -- ~&~ -- ~&~ -- ~\\
10.0 & 4000 &~ -- ~&~ -- ~&~ 0.036 ~&~ 0.830 ~&~ 0.029 ~&~ 1.176
     ~&~ 0.025 ~&~ 1.304 ~\\
11.25 & 1372 &~ 0.0073 ~&~ 0.119 ~&~ 0.0045 ~&~ 0.230 ~&~ 0.0025 ~&~ 0.383
      ~&~ 0.0018 ~&~ 0.426 ~\\
12.5  & 1372 &~ 0.0050 ~&~ 0.085 ~&~ 0.0031 ~&~ 0.172 ~&~ 0.0017 ~&~ 0.287
      ~&~ 0.0012 ~&~ 0.329 ~\\
\hline\hline
\end{tabular}
\footnotetext[1]{108, 256, 500, 1372, 2048}
\footnotetext[2]{108, 256, 500, 864, 1372}
\footnotetext[3]{256, 500, 864, 1372}
\footnotetext[4]{500, 864, 1372}
\footnotetext[5]{864, 1372, 2048}
\footnotetext[6]{500, 864, 1372, 2048, 2916, 4000}
\footnotetext[7]{1372, 2048, 2916, 4000}
\end{center}
\end{table*}

\endgroup

\bibliographystyle{apsrev}
\bibliography{/home/dmitry/p/bib/chem_abbr,/home/dmitry/p/bib/liquids,/home/dmitry/p/bib/dm,/home/dmitry/p/bib/dielectric,/home/dmitry/p/bib/protein,/home/dmitry/p/bib/solvation,/home/dmitry/p/bib/simulations,/home/dmitry/p/bib/etnonlin}

\begin{thebibliography}{43}
\expandafter\ifx\csname natexlab\endcsname\relax\def\natexlab#1{#1}\fi
\expandafter\ifx\csname bibnamefont\endcsname\relax
  \def\bibnamefont#1{#1}\fi
\expandafter\ifx\csname bibfnamefont\endcsname\relax
  \def\bibfnamefont#1{#1}\fi
\expandafter\ifx\csname citenamefont\endcsname\relax
  \def\citenamefont#1{#1}\fi
\expandafter\ifx\csname url\endcsname\relax
  \def\url#1{\texttt{#1}}\fi
\expandafter\ifx\csname urlprefix\endcsname\relax\def\urlprefix{URL }\fi
\providecommand{\bibinfo}[2]{#2}
\providecommand{\eprint}[2][]{\url{#2}}

\bibitem[{\citenamefont{Beck et~al.}(2006)\citenamefont{Beck, Paulaitis, and
  Pratt}}]{Beck:06}
\bibinfo{author}{\bibfnamefont{T.~L.} \bibnamefont{Beck}},
  \bibinfo{author}{\bibfnamefont{M.~E.} \bibnamefont{Paulaitis}},
  \bibnamefont{and} \bibinfo{author}{\bibfnamefont{L.~R.} \bibnamefont{Pratt}},
  \emph{\bibinfo{title}{The potential distribution theorem and models of
  molecular solutions}} (\bibinfo{publisher}{Cambridge University Press},
  \bibinfo{address}{Cambridge}, \bibinfo{year}{2006}).

\bibitem[{\citenamefont{Hansen and McDonald}(2003)}]{Hansen:03}
\bibinfo{author}{\bibfnamefont{J.~P.} \bibnamefont{Hansen}} \bibnamefont{and}
  \bibinfo{author}{\bibfnamefont{I.~R.} \bibnamefont{McDonald}},
  \emph{\bibinfo{title}{Theory of Simple Liquids}}
  (\bibinfo{publisher}{Academic Press}, \bibinfo{address}{Amsterdam},
  \bibinfo{year}{2003}).

\bibitem[{\citenamefont{Landau and Lifshitz}(1984)}]{Landau8}
\bibinfo{author}{\bibfnamefont{L.~D.} \bibnamefont{Landau}} \bibnamefont{and}
  \bibinfo{author}{\bibfnamefont{E.~M.} \bibnamefont{Lifshitz}},
  \emph{\bibinfo{title}{Electrodynamics of continuous media}}
  (\bibinfo{publisher}{Pergamon}, \bibinfo{address}{Oxford},
  \bibinfo{year}{1984}).

\bibitem[{\citenamefont{B{{\"o}}ttcher}(1973)}]{Boettcher:73}
\bibinfo{author}{\bibfnamefont{C.~J.~F.} \bibnamefont{B{{\"o}}ttcher}},
  \emph{\bibinfo{title}{Theory of Electric Polarization}},
  vol.~\bibinfo{volume}{1} (\bibinfo{publisher}{Elsevier},
  \bibinfo{address}{Amsterdam}, \bibinfo{year}{1973}).

\bibitem[{\citenamefont{Allen and Tildesley}(1996)}]{Allen:96}
\bibinfo{author}{\bibfnamefont{M.~P.} \bibnamefont{Allen}} \bibnamefont{and}
  \bibinfo{author}{\bibfnamefont{D.~J.} \bibnamefont{Tildesley}},
  \emph{\bibinfo{title}{Computer Simulation of Liquids}}
  (\bibinfo{publisher}{Clarendon Press}, \bibinfo{address}{Oxford},
  \bibinfo{year}{1996}).

\bibitem[{\citenamefont{Born}(1920)}]{Born:20}
\bibinfo{author}{\bibfnamefont{M.}~\bibnamefont{Born}}, \bibinfo{journal}{Z.
  Phys.} \textbf{\bibinfo{volume}{1}}, \bibinfo{pages}{45}
  (\bibinfo{year}{1920}).

\bibitem[{\citenamefont{Onsager}(1936)}]{Onsager:36}
\bibinfo{author}{\bibfnamefont{L.}~\bibnamefont{Onsager}}, \bibinfo{journal}{J.
  Am. Chem. Soc.} \textbf{\bibinfo{volume}{58}}, \bibinfo{pages}{1486}
  (\bibinfo{year}{1936}).

\bibitem[{\citenamefont{Marenich et~al.}(2008)\citenamefont{Marenich, Cramer,
  and Truhlar}}]{Marenich:08}
\bibinfo{author}{\bibfnamefont{A.~V.} \bibnamefont{Marenich}},
  \bibinfo{author}{\bibfnamefont{C.~J.} \bibnamefont{Cramer}},
  \bibnamefont{and} \bibinfo{author}{\bibfnamefont{D.~G.}
  \bibnamefont{Truhlar}}, \bibinfo{journal}{J.\ Chem.\ Theory Computat.}
  \textbf{\bibinfo{volume}{4}}, \bibinfo{pages}{877} (\bibinfo{year}{2008}).

\bibitem[{\citenamefont{Sokhan and Tildesley}(1997)}]{Sokhan:97}
\bibinfo{author}{\bibfnamefont{V.~P.} \bibnamefont{Sokhan}} \bibnamefont{and}
  \bibinfo{author}{\bibfnamefont{D.~J.} \bibnamefont{Tildesley}},
  \bibinfo{journal}{Mol. Phys.} \textbf{\bibinfo{volume}{92}},
  \bibinfo{pages}{625} (\bibinfo{year}{1997}).

\bibitem[{\citenamefont{Martin and Matyushov}(2008)}]{DMepl:08}
\bibinfo{author}{\bibfnamefont{D.~R.} \bibnamefont{Martin}} \bibnamefont{and}
  \bibinfo{author}{\bibfnamefont{D.~V.} \bibnamefont{Matyushov}},
  \bibinfo{journal}{Europhys.\ Lett.} \textbf{\bibinfo{volume}{82}},
  \bibinfo{pages}{16003} (\bibinfo{year}{2008}).

\bibitem[{\citenamefont{Chandler}(2005)}]{ChandlerNature:05}
\bibinfo{author}{\bibfnamefont{D.}~\bibnamefont{Chandler}},
  \bibinfo{journal}{Nature} \textbf{\bibinfo{volume}{437}},
  \bibinfo{pages}{640} (\bibinfo{year}{2005}).

\bibitem[{\citenamefont{Rashin}(1993)}]{Rashin:93}
\bibinfo{author}{\bibfnamefont{A.~A.} \bibnamefont{Rashin}},
  \bibinfo{journal}{Prog. Biophys. Molec. Biol.} \textbf{\bibinfo{volume}{60}},
  \bibinfo{pages}{73} (\bibinfo{year}{1993}).

\bibitem[{\citenamefont{Lum et~al.}(1999)\citenamefont{Lum, Chandler, and
  Weeks}}]{Lum:99}
\bibinfo{author}{\bibfnamefont{K.}~\bibnamefont{Lum}},
  \bibinfo{author}{\bibfnamefont{D.}~\bibnamefont{Chandler}}, \bibnamefont{and}
  \bibinfo{author}{\bibfnamefont{J.}~\bibnamefont{Weeks}},
  \bibinfo{journal}{J.\ Phys.\ Chem.\ B} \textbf{\bibinfo{volume}{103}},
  \bibinfo{pages}{4570} (\bibinfo{year}{1999}).

\bibitem[{\citenamefont{Hummer and Garde}(1998)}]{HummerPRL:98}
\bibinfo{author}{\bibfnamefont{G.}~\bibnamefont{Hummer}} \bibnamefont{and}
  \bibinfo{author}{\bibfnamefont{S.}~\bibnamefont{Garde}},
  \bibinfo{journal}{Phys. Rev. Lett.} \textbf{\bibinfo{volume}{80}},
  \bibinfo{pages}{4193} (\bibinfo{year}{1998}).

\bibitem[{\citenamefont{Huang and Chandler}(2000)}]{Huang:00}
\bibinfo{author}{\bibfnamefont{D.~M.} \bibnamefont{Huang}} \bibnamefont{and}
  \bibinfo{author}{\bibfnamefont{D.}~\bibnamefont{Chandler}},
  \bibinfo{journal}{Phys. Rev. E} \textbf{\bibinfo{volume}{61}},
  \bibinfo{pages}{1501} (\bibinfo{year}{2000}).

\bibitem[{\citenamefont{Ashbaugh and Pratt}(2006)}]{Ashbaugh:06}
\bibinfo{author}{\bibfnamefont{H.~S.} \bibnamefont{Ashbaugh}} \bibnamefont{and}
  \bibinfo{author}{\bibfnamefont{L.~R.} \bibnamefont{Pratt}},
  \bibinfo{journal}{Rev.\ Mod.\ Phys.} \textbf{\bibinfo{volume}{78}},
  \bibinfo{eid}{159} (\bibinfo{year}{2006}).

\bibitem[{\citenamefont{Maggs and Everaers}(2006)}]{Maggs:06}
\bibinfo{author}{\bibfnamefont{A.~C.} \bibnamefont{Maggs}} \bibnamefont{and}
  \bibinfo{author}{\bibfnamefont{R.}~\bibnamefont{Everaers}},
  \bibinfo{journal}{Phys. Rev. Lett.} \textbf{\bibinfo{volume}{96}},
  \bibinfo{pages}{230603} (\bibinfo{year}{2006}).

\bibitem[{\citenamefont{Rudas et~al.}(2006)\citenamefont{Rudas, Schr{\"o}der,
  and Steinhauser}}]{Rudas:06}
\bibinfo{author}{\bibfnamefont{T.}~\bibnamefont{Rudas}},
  \bibinfo{author}{\bibfnamefont{C.}~\bibnamefont{Schr{\"o}der}},
  \bibnamefont{and}
  \bibinfo{author}{\bibfnamefont{O.}~\bibnamefont{Steinhauser}},
  \bibinfo{journal}{J. Chem. Phys.} \textbf{\bibinfo{volume}{124}},
  \bibinfo{pages}{234908} (\bibinfo{year}{2006}).

\bibitem[{\citenamefont{Kuharski et~al.}(1988)\citenamefont{Kuharski, Bader,
  Chandler, Sprik, Klein, and Impey}}]{Kuharski:88}
\bibinfo{author}{\bibfnamefont{R.~A.} \bibnamefont{Kuharski}},
  \bibinfo{author}{\bibfnamefont{J.~S.} \bibnamefont{Bader}},
  \bibinfo{author}{\bibfnamefont{D.}~\bibnamefont{Chandler}},
  \bibinfo{author}{\bibfnamefont{M.}~\bibnamefont{Sprik}},
  \bibinfo{author}{\bibfnamefont{M.~L.} \bibnamefont{Klein}}, \bibnamefont{and}
  \bibinfo{author}{\bibfnamefont{R.~W.} \bibnamefont{Impey}},
  \bibinfo{journal}{J. Chem. Phys.} \textbf{\bibinfo{volume}{89}},
  \bibinfo{pages}{3248} (\bibinfo{year}{1988}).

\bibitem[{\citenamefont{Blumberger and Sprik}(2005)}]{Blumberger:05}
\bibinfo{author}{\bibfnamefont{J.}~\bibnamefont{Blumberger}} \bibnamefont{and}
  \bibinfo{author}{\bibfnamefont{M.}~\bibnamefont{Sprik}}, \bibinfo{journal}{J.
  Phys. Chem. B} \textbf{\bibinfo{volume}{109}}, \bibinfo{pages}{6793}
  (\bibinfo{year}{2005}).

\bibitem[{\citenamefont{Aqvist and Hansson}(1996)}]{Aqvist:96}
\bibinfo{author}{\bibfnamefont{J.}~\bibnamefont{Aqvist}} \bibnamefont{and}
  \bibinfo{author}{\bibfnamefont{T.}~\bibnamefont{Hansson}},
  \bibinfo{journal}{J. Phys. Chem.} \textbf{\bibinfo{volume}{100}},
  \bibinfo{pages}{9512} (\bibinfo{year}{1996}).

\bibitem[{\citenamefont{Ashbaugh}(2000)}]{Ashbaugh:00}
\bibinfo{author}{\bibfnamefont{H.~S.} \bibnamefont{Ashbaugh}},
  \bibinfo{journal}{J. Phys. Chem. B} \textbf{\bibinfo{volume}{104}},
  \bibinfo{pages}{7235} (\bibinfo{year}{2000}).

\bibitem[{\citenamefont{Cerutti et~al.}(2007)\citenamefont{Cerutti, Baker, and
  McCammon}}]{Cerutti:07}
\bibinfo{author}{\bibfnamefont{D.~S.} \bibnamefont{Cerutti}},
  \bibinfo{author}{\bibfnamefont{N.~A.} \bibnamefont{Baker}}, \bibnamefont{and}
  \bibinfo{author}{\bibfnamefont{J.~A.} \bibnamefont{McCammon}},
  \bibinfo{journal}{J.\ Chem.\ Phys.} \textbf{\bibinfo{volume}{127}},
  \bibinfo{eid}{155101} (\bibinfo{year}{2007}).

\bibitem[{\citenamefont{Yu and Karplus}(1988)}]{Yu:88}
\bibinfo{author}{\bibfnamefont{H.-A.} \bibnamefont{Yu}} \bibnamefont{and}
  \bibinfo{author}{\bibfnamefont{M.}~\bibnamefont{Karplus}},
  \bibinfo{journal}{J. Chem. Phys.} \textbf{\bibinfo{volume}{89}},
  \bibinfo{pages}{2366} (\bibinfo{year}{1988}).

\bibitem[{\citenamefont{Ben-Amotz et~al.}(2005)\citenamefont{Ben-Amotz,
  Raineri, and Stell}}]{Ben-Amotz:05}
\bibinfo{author}{\bibfnamefont{D.}~\bibnamefont{Ben-Amotz}},
  \bibinfo{author}{\bibfnamefont{F.~O.} \bibnamefont{Raineri}},
  \bibnamefont{and} \bibinfo{author}{\bibfnamefont{G.}~\bibnamefont{Stell}},
  \bibinfo{journal}{J. Phys. Chem. B} \textbf{\bibinfo{volume}{109}},
  \bibinfo{pages}{6866} (\bibinfo{year}{2005}).

\bibitem[{\citenamefont{Neumann}(1986)}]{Neumann:86}
\bibinfo{author}{\bibfnamefont{M.}~\bibnamefont{Neumann}},
  \bibinfo{journal}{Mol. Phys.} \textbf{\bibinfo{volume}{57}},
  \bibinfo{pages}{97} (\bibinfo{year}{1986}).

\bibitem[{\citenamefont{Matyushov and Ladanyi}(1999)}]{DMjcp1:99}
\bibinfo{author}{\bibfnamefont{D.~V.} \bibnamefont{Matyushov}}
  \bibnamefont{and} \bibinfo{author}{\bibfnamefont{B.~M.}
  \bibnamefont{Ladanyi}}, \bibinfo{journal}{J. Chem. Phys.}
  \textbf{\bibinfo{volume}{110}}, \bibinfo{pages}{994} (\bibinfo{year}{1999}).

\bibitem[{\citenamefont{Chan et~al.}(1979)\citenamefont{Chan, Mitchell, and
  Ninham}}]{Chan:79}
\bibinfo{author}{\bibfnamefont{D.~Y.~C.} \bibnamefont{Chan}},
  \bibinfo{author}{\bibfnamefont{D.~J.} \bibnamefont{Mitchell}},
  \bibnamefont{and} \bibinfo{author}{\bibfnamefont{B.~W.}
  \bibnamefont{Ninham}}, \bibinfo{journal}{J. Chem. Phys.}
  \textbf{\bibinfo{volume}{70}}, \bibinfo{pages}{2946} (\bibinfo{year}{1979}).

\bibitem[{\citenamefont{Wertheim}(1971)}]{Wertheim:71}
\bibinfo{author}{\bibfnamefont{M.~S.} \bibnamefont{Wertheim}},
  \bibinfo{journal}{J. Chem. Phys.} \textbf{\bibinfo{volume}{55}},
  \bibinfo{pages}{4291} (\bibinfo{year}{1971}).

\bibitem[{\citenamefont{Freasier and Isbister}(1979)}]{Freasier:79}
\bibinfo{author}{\bibfnamefont{B.~C.} \bibnamefont{Freasier}} \bibnamefont{and}
  \bibinfo{author}{\bibfnamefont{D.~J.} \bibnamefont{Isbister}},
  \bibinfo{journal}{Mol. Phys.} \textbf{\bibinfo{volume}{38}},
  \bibinfo{pages}{81} (\bibinfo{year}{1979}).

\bibitem[{\citenamefont{Larsen et~al.}(1977)\citenamefont{Larsen, Rasaiah, and
  Stell}}]{Larsen:77}
\bibinfo{author}{\bibfnamefont{B.}~\bibnamefont{Larsen}},
  \bibinfo{author}{\bibfnamefont{J.~C.} \bibnamefont{Rasaiah}},
  \bibnamefont{and} \bibinfo{author}{\bibfnamefont{G.}~\bibnamefont{Stell}},
  \bibinfo{journal}{Mol. Phys.} \textbf{\bibinfo{volume}{33}},
  \bibinfo{pages}{987} (\bibinfo{year}{1977}).

\bibitem[{\citenamefont{Gupta and Matyushov}(2004)}]{DMjpca:04}
\bibinfo{author}{\bibfnamefont{S.}~\bibnamefont{Gupta}} \bibnamefont{and}
  \bibinfo{author}{\bibfnamefont{D.~V.} \bibnamefont{Matyushov}},
  \bibinfo{journal}{J. Phys. Chem. A} \textbf{\bibinfo{volume}{108}},
  \bibinfo{pages}{2087} (\bibinfo{year}{2004}).

\bibitem[{\citenamefont{Matyushov and Ladanyi}(1997)}]{DMjcp3:97}
\bibinfo{author}{\bibfnamefont{D.~V.} \bibnamefont{Matyushov}}
  \bibnamefont{and} \bibinfo{author}{\bibfnamefont{B.~M.}
  \bibnamefont{Ladanyi}}, \bibinfo{journal}{J. Chem. Phys.}
  \textbf{\bibinfo{volume}{107}}, \bibinfo{pages}{5815} (\bibinfo{year}{1997}).

\bibitem[{\citenamefont{Weis}(2002)}]{Weis:02}
\bibinfo{author}{\bibfnamefont{J.~J.} \bibnamefont{Weis}},
  \bibinfo{journal}{Mol. Phys.} \textbf{\bibinfo{volume}{100}},
  \bibinfo{pages}{579} (\bibinfo{year}{2002}).

\bibitem[{\citenamefont{Rajamani et~al.}(2004)\citenamefont{Rajamani, Ghosh,
  and Garde}}]{Rajamani:04}
\bibinfo{author}{\bibfnamefont{S.}~\bibnamefont{Rajamani}},
  \bibinfo{author}{\bibfnamefont{T.}~\bibnamefont{Ghosh}}, \bibnamefont{and}
  \bibinfo{author}{\bibfnamefont{S.}~\bibnamefont{Garde}},
  \bibinfo{journal}{J.\ Chem.\ Phys.} \textbf{\bibinfo{volume}{120}},
  \bibinfo{pages}{4457} (\bibinfo{year}{2004}).

\bibitem[{\citenamefont{Frodl and Dietrich}(1992)}]{Frodl:92}
\bibinfo{author}{\bibfnamefont{P.}~\bibnamefont{Frodl}} \bibnamefont{and}
  \bibinfo{author}{\bibfnamefont{S.}~\bibnamefont{Dietrich}},
  \bibinfo{journal}{Phys. Rev. A} \textbf{\bibinfo{volume}{45}},
  \bibinfo{pages}{7330} (\bibinfo{year}{1992}).

\bibitem[{\citenamefont{DeLeeuw et~al.}(1980)\citenamefont{DeLeeuw, Perram, and
  Smith}}]{Leeuw:80}
\bibinfo{author}{\bibfnamefont{S.~W.} \bibnamefont{DeLeeuw}},
  \bibinfo{author}{\bibfnamefont{J.~W.} \bibnamefont{Perram}},
  \bibnamefont{and} \bibinfo{author}{\bibfnamefont{E.~R.} \bibnamefont{Smith}},
  \bibinfo{journal}{Proc. R. Soc. London Ser. A}
  \textbf{\bibinfo{volume}{373}}, \bibinfo{pages}{27} (\bibinfo{year}{1980}).

\bibitem[{\citenamefont{Hummer et~al.}(1998)\citenamefont{Hummer, Pratt, and
  Garcia}}]{Hummer:96}
\bibinfo{author}{\bibfnamefont{G.}~\bibnamefont{Hummer}},
  \bibinfo{author}{\bibfnamefont{L.~R.} \bibnamefont{Pratt}}, \bibnamefont{and}
  \bibinfo{author}{\bibfnamefont{A.~E.} \bibnamefont{Garcia}},
  \bibinfo{journal}{J. Phys. Chem. A} \textbf{\bibinfo{volume}{102}},
  \bibinfo{pages}{7885} (\bibinfo{year}{1998}).

\bibitem[{\citenamefont{Milischuk and Matyushov}(2002)}]{DMjpca:02}
\bibinfo{author}{\bibfnamefont{A.}~\bibnamefont{Milischuk}} \bibnamefont{and}
  \bibinfo{author}{\bibfnamefont{D.~V.} \bibnamefont{Matyushov}},
  \bibinfo{journal}{J. Phys. Chem. A} \textbf{\bibinfo{volume}{106}},
  \bibinfo{pages}{2146} (\bibinfo{year}{2002}).

\bibitem[{\citenamefont{Matyushov}(2005)}]{DMjcp1:05}
\bibinfo{author}{\bibfnamefont{D.~V.} \bibnamefont{Matyushov}},
  \bibinfo{journal}{J. Chem. Phys.} \textbf{\bibinfo{volume}{122}},
  \bibinfo{pages}{044502} (\bibinfo{year}{2005}).

\bibitem[{\citenamefont{Ghorai and Matyushov}(2006)}]{DMjpca1:06}
\bibinfo{author}{\bibfnamefont{P.~K.} \bibnamefont{Ghorai}} \bibnamefont{and}
  \bibinfo{author}{\bibfnamefont{D.~V.} \bibnamefont{Matyushov}},
  \bibinfo{journal}{J. Phys. Chem. A} \textbf{\bibinfo{volume}{110}},
  \bibinfo{pages}{8857} (\bibinfo{year}{2006}).

\bibitem[{\citenamefont{Ben-Amotz and Underwood}(2008)}]{Ben-AmotzAcc:08}
\bibinfo{author}{\bibfnamefont{D.}~\bibnamefont{Ben-Amotz}} \bibnamefont{and}
  \bibinfo{author}{\bibfnamefont{R.}~\bibnamefont{Underwood}},
  \bibinfo{journal}{Acc.\ Chem.\ Res.} p. \bibinfo{pages}{to be published}
  (\bibinfo{year}{2008}).

\bibitem[{\citenamefont{LeBard and Matyushov}(2008)}]{DMjcp2:08}
\bibinfo{author}{\bibfnamefont{D.~N.} \bibnamefont{LeBard}} \bibnamefont{and}
  \bibinfo{author}{\bibfnamefont{D.~V.} \bibnamefont{Matyushov}},
  \bibinfo{journal}{J. Chem. Phys.} \textbf{\bibinfo{volume}{128}},
  \bibinfo{pages}{155106} (\bibinfo{year}{2008}).

\end{thebibliography}

\end{document}